\documentclass[times,twocolumn,final]{elsarticle}
\usepackage{framed,multirow}
\usepackage{amssymb}
\usepackage{latexsym}
\usepackage{url}
\usepackage{xcolor}
\usepackage{booktabs}
\usepackage{hyperref}
\usepackage{soul}
\usepackage{geometry}
\usepackage{flushend}
\usepackage{graphicx}
\usepackage{subfigure}
\usepackage{booktabs} 
\usepackage{cite}
\usepackage{array, caption, threeparttable}
\usepackage[font=small,labelfont=bf,labelsep=none]{caption}
\usepackage{natbib}
\usepackage{amsmath,amssymb,amsfonts}
\usepackage{algorithmic}
\usepackage{textcomp}
\usepackage{booktabs}
\usepackage{xspace}
\usepackage{svg}
\usepackage{caption}
\usepackage{footnote}
\usepackage{makecell}
\usepackage{multirow}
\usepackage{enumitem}
\usepackage{array}
\usepackage[nolist,nohyperlinks]{acronym}
\usepackage{tabularx}
\usepackage{float}
\usepackage{balance}
\usepackage[acronym]{glossaries}
\makeglossaries
\makeatletter

\newcommand{\Rmnum}[1]{\expandafter\@slowromancap\romannumeral #1@}
\makeatother
\definecolor{newcolor}{rgb}{.8,.349,.1}
\journal{}

\begin{document}

\begin{frontmatter}

\title{WAL-Net: Weakly supervised auxiliary task learning network for carotid plaques classification}%
\author[1]{Haitao Gan \corref{cor1}}
\cortext[cor1]{Corresponding authors}\ead{htgan01@hbut.edu.cn}
\author[1]{Lingchao Fu}
\author[1]{Ran Zhou}
\author[1]{Weiyan Gan}
\author[2]{Furong Wang}
\author[3]{Xiaoyan Wu}
\author[1]{Zhi Yang}
\author[1]{Zhongwei Huang}

\address[1]{School of Computer Science, Hubei University of Technology, Wuhan, 430068, China}
\address[2]{Liyuan Hospital, Tongji Medical College, Huazhong University of Science and Technology, Wuhan, China}
\address[3]{Department of Cardiology, Zhongnan Hospital, Wuhan University, Wuhan, China}

\begin{abstract}
The classification of carotid artery ultrasound images is a crucial means for diagnosing carotid plaques, holding significant clinical relevance for predicting the risk of stroke. Recent research suggests that utilizing plaque segmentation as an auxiliary task for classification can enhance performance by leveraging the correlation between segmentation and classification tasks. However, this approach relies on obtaining a substantial amount of challenging-to-acquire segmentation annotations. This paper proposes a novel weakly supervised auxiliary task learning network model (WAL-Net) to explore the interdependence between carotid plaque classification and segmentation tasks. The plaque classification task is primary task, while the plaque segmentation task serves as an auxiliary task, providing valuable information to enhance the performance of the primary task. Weakly supervised learning is adopted in the auxiliary task to completely break away from the dependence on segmentation annotations. Experiments and evaluations are conducted on a dataset comprising 1270 carotid plaque ultrasound images from Wuhan University Zhongnan Hospital. Results indicate that the proposed method achieved an approximately 1.3\% improvement in carotid plaque classification accuracy compared to the baseline network. Specifically, the accuracy of mixed-echoic plaques classification increased by approximately 3.3\%, demonstrating the effectiveness of our approach.
\end{abstract}

\end{frontmatter}

\section{Introduction}
{
Ischemic stroke stands as a primary cause of disability and mortality among cardiovascular disease patients globally ~\citep{beaglehole2008global}. Atherosclerosis constitutes the predominant pathological process leading to the majority of ischemic strokes, with carotid artery plaques being a manifestation of atherosclerosis. The rupture and detachment of carotid artery plaques contribute to thrombosis or vascular stenosis, thereby precipitating ischemic stroke events. In the year 2020, approximately 21.1\% of the global population aged 30 to 79 exhibited carotid artery plaques ~\citep{song2020global}. Based on distinct ultrasound echo characteristics, carotid artery plaques can be categorized into three types: hypoechoic plaques, hyperechoic plaques, and mixed-echoic plaques ~\citep{aburahma2002carotid} ~\citep{olender2022impact}. Among these plaques, hyperechoic plaques are relatively stable, while the remaining two types are prone to rupture, leading to ischemic stroke. Consequently, the study, identification, and treatment of carotid artery plaque categories are of paramount importance. Ultrasound examination at peripheral arteries and carotid arteries is currently the preferred non-invasive method for carotid artery assessment. Widely employed for screening and follow-up of carotid artery atherosclerotic lesions, ultrasound examination reveals the location and size of plaques, as well as the site and severity of luminal stenosis. However, this process demands high levels of concentration from medical professionals, leading to the potential for misdiagnosis and incurring significant time costs. The utilization of deep learning for auxiliary diagnosis emerges as a viable solution, addressing the aforementioned issues while also offering potential enhancements in diagnostic accuracy and efficiency.

The segmentation and classification tasks of carotid artery plaque ultrasound images are two key components in deep learning-based processing of carotid artery plaque ultrasound images. The purpose of the classification task is to diagnose the exact category of carotid artery plaques (e.g.,  hypoechoic plaques or hyperechoic plaques), while the segmentation task is employed to detect the precise location and shape of carotid artery plaques. In practice, there exists a certain degree of correlation between the segmentation and classification tasks of carotid artery plaques. For instance, the output of the segmentation task can be utilized to enhance the weight of lesion areas in the features for the classification task, thereby improving classification performance. Currently, there are two main approaches for leveraging segmentation tasks to enhance classification task performance: non-end-to-end methods and end-to-end methods. Non-end-to-end methods involve training two or more separate models to perform segmentation and classification tasks independently. The output of one model is then used to enhance the output of another model. In the field of medical image processing, various methods of this kind have been proposed. For example, Miao Wang et al. introduced a parallel polyp segmentation and classification method to explore the correlation between the two tasks ~\citep{wang2023efficient}. This method utilizes the preliminary segmentation of samples as an additional channel input to the classification network, enhancing the classification performance. Amirreza Mahbod and colleagues investigated the impact of different approaches to handling segmentation tasks on classification tasks in the context of skin disease image processing ~\citep{mahbod2020effects}.In contrast, end-to-end methods design a unified network model capable of simultaneously executing different tasks through multi-task learning. These methods typically share parameters and loss functions among different tasks to learn common and useful information. For instance, He et al. proposed the Lesion Area Extraction (LAE) Module, which employs an expansive lesion area cropping strategy to filter background noise from classification features, thus improving classification performance ~\citep{he2023joint}. Ou et al. introduced a multi-task network model that simultaneously performs carotid artery plaque classification and segmentation ~\citep{ou2022auxiliary}. In this approach, the segmentation task identifies the pathological areas of carotid artery plaques, and then the weights of these areas in the features for the classification task are reinforced to improve classification performance.In summary, non-end-to-end methods may face the challenge of high training costs as they require training additional models for different tasks. On the other hand, end-to-end methods need to better exploit the correlation between classification and segmentation tasks to avoid the features learned being overly negatively influenced by different task objectives ~\citep{zhang2021survey}.

While designing network models to perform multiple tasks can leverage segmentation tasks to enhance the effectiveness of classification tasks, obtaining segmentation labels for carotid artery plaque ultrasound images is challenging and requires a significant time investment from experts or medical professionals. In our previous work, we addressed this challenge by employing semi-supervised learning for the segmentation task of carotid artery plaque ultrasound images, aiming to reduce dependence on segmentation annotations ~\citep{fu2023smc}. This approach yielded promising results. However, considering our primary focus on the performance of the network model in the main task, namely carotid artery plaque classification, it is essential to treat the segmentation task as a purer auxiliary task. For example, employing weakly supervised learning for the segmentation task can be explored, allowing segmentation task to be performed even in the absence of segmentation annotations.

Based on the considerations mentioned above, this paper proposes a novel end-to-end multi-task learning network model for carotid artery plaque classification, named the Weakly supervised Auxiliary Learning Network (WAL-Net). WAL-Net introduces an auxiliary task, namely weakly supervised segmentation, for the primary task of carotid artery plaque classification. The purpose of the auxiliary task is to identify the specific location and shape of the pathological regions of carotid artery plaques in ultrasound images. WAL-Net comprises a shared encoder, a decoder for segmentation, and a classification task head.The weakly supervised segmentation task is supervised by the information generated from the proposed Pseudo mask Generation Module (PGM), where the localization information is obtained through attention methods, and affinity is guided by superpixel method. The shared encoder is responsible for extracting multidimensional features of carotid artery plaques. Additionally, WAL-Net incorporates a Region of Interest cropping Module (RCM), utilizing the output predictions from the segmentation decoder to obtain the location information of the lesion plaques. This information is then used to enhance the classification features in the encoder. The enhanced classification features are input into the classification task head to improve the performance of the classification task. The overall architecture of WAL-Net is designed to seamlessly integrate the weakly supervised segmentation task as a purer auxiliary task, with the shared encoder efficiently capturing relevant features for both segmentation and classification tasks.

In summary, our work entails the following contributions:

\begin{itemize}
\item WAL-Net seamlessly integrates supervised classification tasks with weakly supervised segmentation tasks into a unified multi-task learning model. In comparison to non-end-to-end methodologies, WAL-Net is proficient in concurrently executing classification and segmentation tasks.

\item Our proposed Pseudo mask Generation Module (PGM) amalgamates attention methods with superpixel methods to generate pseudo-masks. Employing weakly supervised learning enhances segmentation outcomes, eliminating the need for reliance on segmentation annotations.

\item Our proposed Region of Interest cropping Module (RCM) adaptively acquires Regions of Interest (ROI) from features of varying scales and subsequently enhances them. This explicit utilization of inter-task correlations in multi-task learning proves to be instrumental.

\item Experimental results demonstrate that WAL-Net achieves superior performance in carotid artery plaque classification. Furthermore, a ablation study corroborates the efficacy of each module within the proposed framework.
\end{itemize}

The remaining structure of the manuscript is as follows: After an introducing carotid artery plaques in the medical domain, Section 2 discusses recent investigations in this field. Subsequently, Section 3 delves into the materials and methods employed in the manuscript. Section 4 presents the results of the research findings, and finally, Section 5 provides a summary of the study.

\section{Related Work}
{
To date, numerous researchers have proposed various methods for the classification of carotid artery plaques. In conventional classification approaches, Cevlan et al.~\citep{ceylan2007classification} employed Principal Component Analysis (PCA) and Complex-Valued Artificial Neural Network (CVANN) for the classification of carotid artery Doppler ultrasound signals. Tsiaparas et al. ~\citep{tsiaparas2012assessment} utilized Support Vector Machines (SVM) to assess the capability of three multiscale transformation methods in extracting features related to carotid artery atherosclerotic plaques. Chaudhry et al. ~\citep{chaudhry2013automatic} introduced a technique that employs SVM and intima-media thickness as a feature vector for the classification of carotid artery ultrasound images. However, these traditional machine learning methods predominantly rely on one or more handcrafted features, rendering them incapable of accurately and comprehensively describing the state of carotid artery plaques.

Auxiliary task learning is a form of multi-task learning (MTL) ~\citep{ruder2017overview} aimed at benefiting from multiple tasks by incorporating suitable auxiliary tasks. Auxiliary task learning exhibits superior generalization characteristics compared to single-task learning. For instance, Zhang et al.  ~\citep{zhang2015learning} proposed jointly learning facial fine-grained features and head pose features in a network model, treating the learning task of head pose features as an auxiliary task to address the challenges of facial image feature learning under conditions such as image occlusion or pose variations. He et al. ~\citep{he2023joint} introduced a method for lesion segmentation and classification in skin disease images, along with edge segmentation, to explore correlations among multiple tasks. The edge segmentation task in skin disease images was utilized as an auxiliary task to leverage edge information and enhance the features related to edges in the image segmentation task. Liebel et al. ~\citep{liebel2018auxiliary}, in their study on auxiliary task learning, observed that the selection of auxiliary tasks should involve tasks that are easy to learn and obtain annotations. Appropriately chosen seemingly unrelated auxiliary tasks can significantly enhance the performance of the primary task.

The weakly supervised segmentation, as an application within weakly supervised learning ~\citep{zhou2018brief}, aims to enhance training effectiveness in scenarios with limited supervisory information or reduce dependence on such supervision. The weakly supervised segmentation adopted in this paper is based on image-level labels. Despite numerous studies in recent years on weakly supervised segmentation using image-level labels, for instance, Yuliang Zou et al. ~\citep{zou2020pseudoseg} utilized characteristics from semi-supervised learning, merging pseudo-labels generated from diverse sources and various data augmentations to improve the effectiveness of weakly supervised segmentation. However, since this paper necessitates leveraging weakly supervised auxiliary tasks to enhance the primary task's performance, most current non-end-to-end weakly supervised learning methods are unsuitable. Approaches based on Class Activation Maps (CAM) require initial backward of classification task losses before implementation. For example, Sun et al. ~\citep{sun2021ecs} proposed improving CAM performance using erased images to obtain more accurate pseudo-masks, a method contradictory to end-to-end auxiliary task learning. Another approach involves employing attention mechanisms to guide weakly supervised segmentation. Kunpeng Li et al. ~\citep{li2018tell}, to address the inability to employ end-to-end learning in weakly supervised segmentation, proposed using attention maps as priors for feature localization and semantic segmentation tasks. This method extracts localization and segmentation information from attention maps without the need for additional segmentation labels. Sheng Yi et al. ~\citep{yi2022weakly} employed superpixels to guide semantic affinity between pixels, amalgamating superpixel and localization information. This method not only focuses on localization information by CAM but also considers appearance information based on superpixel method.
}

\begin{figure*}[!t]
	\centering
	\includegraphics[width=0.98\textwidth]{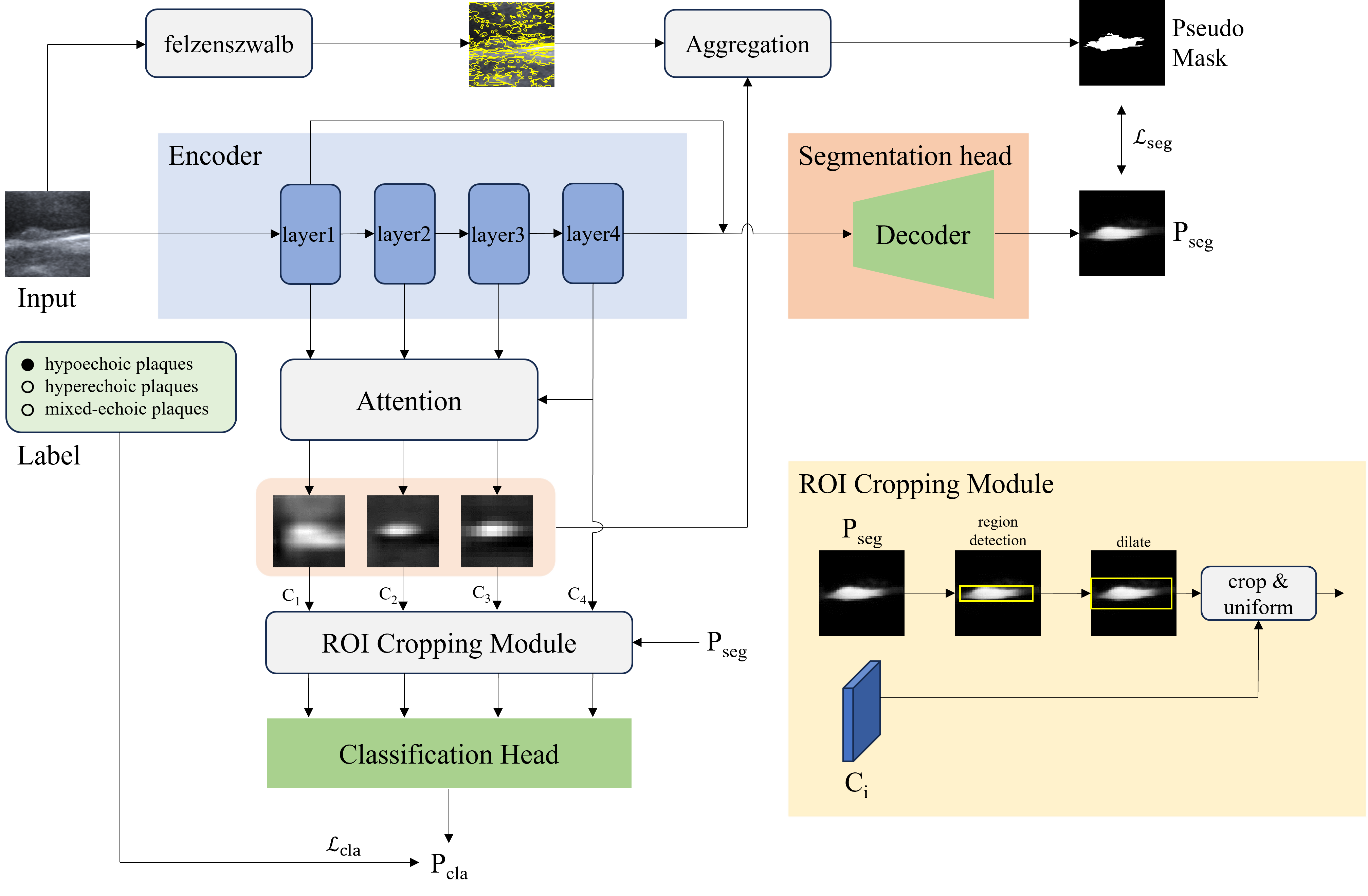}
	\caption{Overview of the proposed WAL-Net. Bottom right: Structure of the ROI Cropping Module.}
	\label{fig:1}
\end{figure*}

\section{Methodology}
{
WAL-Net comprises a shared encoder and two distinct task heads. In WAL-Net, the weakly supervised segmentation task serves as an auxiliary task with the aim of enhancing the performance of the primary task, namely the classification task. As illustrated in Fig.\ref{fig:1}, preprocessed samples of carotid artery ultrasound images are utilized as inputs, and a shared encoder is employed to extract features at different scales. Subsequently, deep and shallow features are fused in the segmentation decoder (utilizing the decoder method proposed by Deeplabv3+ ~\citep{chen2018encoder}) to obtain corresponding segmentation predictions. The role of segmentation predictions is to assist the execution of the classification task. After obtaining features at different scales, attention gates ~\citep{schlemper2018attention} are employed by the shared encoder to generate attention maps at different scales, enhancing the features. Subsequently, the attention-enriched features and segmentation predictions are jointly fed into the RCM to obtain multi-dimensional features of the amplified region of interest. These features are then input into the classification task head to obtain the final classification prediction. In WAL-Net, the classification and segmentation tasks collaborate explicitly, leveraging certain characteristics of multi-task learning to improve overall performance. Each module will be detailed below.

\subsection{The weakly supervised segmentation task}
{The purpose of the auxiliary task is to learn features that are beneficial for the primary task and provide certain enhancements to the primary task. In our case, we employ the weakly supervised segmentation task on carotid artery ultrasound images as the auxiliary task. The encoder of the segmentation network is shared with the encoder of the primary task. The decoder structure follows deeplabv3+ ~\citep{chen2018encoder}, which combines shallow and deep features to improve the final segmentation prediction. We introduce the PGM to generate pseudo masks, which are used to supervise the segmentation predictions.

 Pseudo mask Generation Module (PGM): The process of generating pseudo masks by the PGM is illustrated in Fig.\ref{fig:2}. In WAL-Net, the input image is initially processed by Felzenszwalb's superpixel segmentation method ~\citep{felzenszwalb2004efficient}, resulting in a superpixel map. The attention maps at different scales, obtained through attention mechanisms in the shared encoder, are then fused and regionally averaged, assigning weighted values to the superpixel map. After binarization, this produces a weighted segmentation map, serving as the supervision for the weakly supervised segmentation task, as depicted in Fig.\ref{fig:2}. The attention method provides localization information for the pseudo mask, while the superpixel method guides the affinity. The module combines attention and superpixel methods to generate the pseudo mask.

\begin{figure*}[!t]
	\centering
	\includegraphics[width=0.98\textwidth]{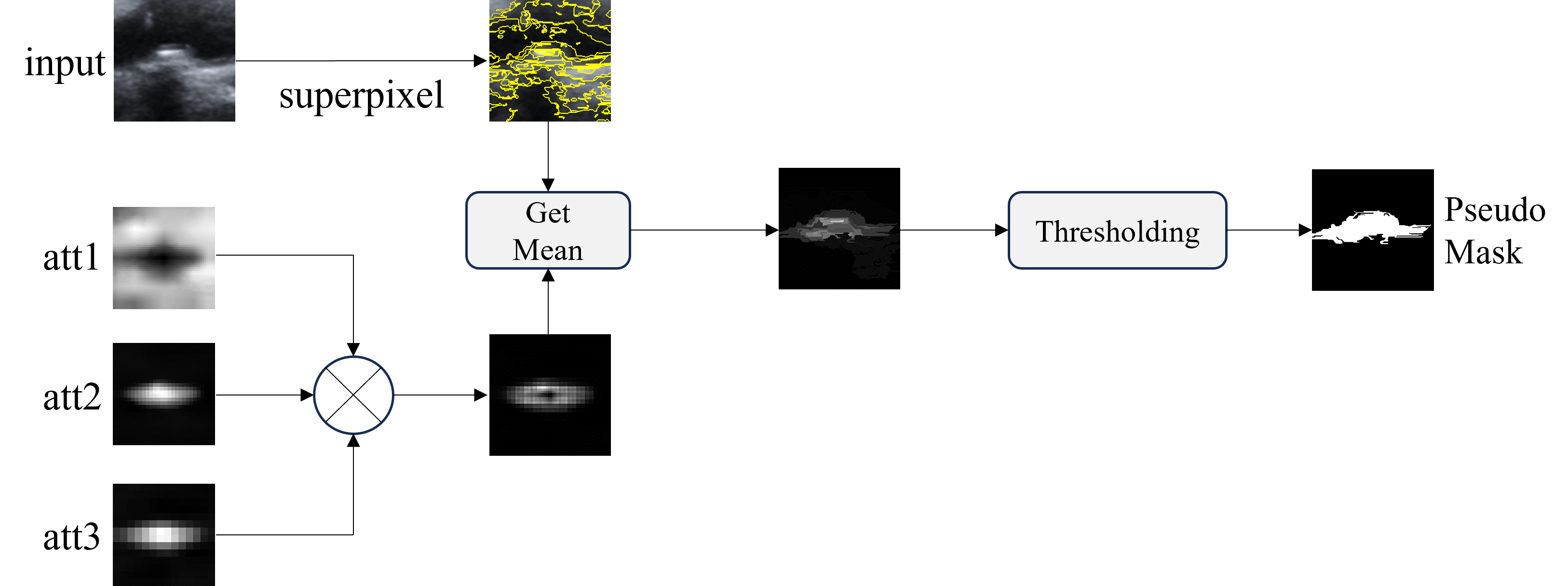}
	\caption{Structure of the Pseudo mask Generation Module.}
	\label{fig:2}
\end{figure*}

The fusion of attention maps at three different hierarchical levels is achieved through element-wise multiplication (as expressed in Eq.\ref{eq:1}). This choice is made because the attention map for shallow features tends to have clearer contours, while the attention map for deep features provides more accurate positioning. Combining attention maps at different levels yields more complete and accurate positional information. After the fusion of attention maps at different levels, averaging is performed within the segmented regions of the superpixel map, resulting in a fused segmentation map, as indicated by Eq.\ref{eq:2}.

\begin{equation}\label{eq:1}
		B_i = (A^1_i *A^2_i * A^3_i)
\end{equation}

Here,  ${i} \in [1,n]$, where $n$ represents the number of training samples in a batch, and $i$ represents the $i$-th sample in a training sample.  ${A^1}$, ${A^2}$, and ${A^3}$ represent the attention maps at three hierarchical levels. The fusion attention map ${B}$ is obtained by multiplying the attention maps from the three levels

\begin{equation}\label{eq:2}
		D_{c=j} = \frac{1}{n}\sum_{k=1}^{n}B_{c=k}
\end{equation}

Here, $C$ represents the superpixel map obtained through Felzenszwalb, $j\in[1,m]$, where $m$ represents the number of segmentation regions in $C$, and $j$ represents the $j$-th segmentation region. $n$ represents the number of pixels in the $j$-th segmentation region of $C$.The fused attention map $B$ is averaged for each corresponding segmentation region in the unsupervised segmentation map $C$. The result is assigned to that segmentation region. After this operation, a pseudo mask $D$ is obtained through binarization, which is then used to supervise the weakly supervised segmentation task. The cross-entropy loss function for this task is formulated as Eq.\ref{eq:3}.

\begin{equation}\label{eq:3}
		loss_{seg} = \frac{1}{L}\sum_{j=1}^{L}\sum_{m=1}^{1*W*H} D_{jm}\log(s_{jm})
\end{equation}

Here, $L$ represents the number of training samples in the dataset, $D$ represents the pseudo mask, and $s$ represents the segmentation prediction made by the network for the training sample. $D_{jm}$ and $s_{jm}$ denote the $m$-th pixel value of the $j$-th sample in $D$ and $s$, respectively.

} 

\subsubsection{The classification task}
{WAL-Net focuses on the primary task of classifying carotid artery plaque ultrasound images, which involves categorizing plaques into three distinct types: hypoechoic plaques, hyperechoic plaques, and mixed-echoic plaques. The classification network employed by WAL-Net bears similarity to the architecture introduced in resnest ~\citep{zhang2022resnest}, allowing it to extract profound features from carotid artery ultrasound images. Furthermore, WAL-Net incorporates Attention Gates ~\citep{schlemper2018attention} into its encoder, as proposed by Jo Schlemper et al. This attention mechanism is well-suited for medical image analysis and provides detailed, visually interpretable attention maps. Recognizing the utility of segmentation predictions obtained during the auxiliary task for the primary task, we introduce the RCM to explicitly enhance the classification task using segmentation predictions.

ROI Cropping Module (RCM): Upon obtaining segmentation predictions for carotid artery plaque ultrasound images during the auxiliary task, both the segmentation predictions and features from different depths are fed into RCM. Firstly, RCM binarizes the segmentation predictions (with a binary threshold set in this paper, values greater than or equal to 0.5 are set to 1, while values less than 0.5 are set to 0). Subsequently, RCM obtains the bounding box of the lesion area in the segmentation predictions. The bounding box is expanded by $\lambda$ in all directions (with $\lambda$ set to 1/7 of the matrix size in this paper) to retain a portion of the normal vascular wall or other background. Finally, the bounding box is resized uniformly and serves as input for the classification task head.In the classification task head, WAL-Net separately feeds features from different depths into fully connected layers, and the average of the results is taken as the final output. The loss function for the classification task is determined by the cross-entropy loss function, which calculates the difference between the predicted lesion classification and the actual lesion type. The loss function is expressed as Eq.\ref{eq:4}.

\begin{equation}\label{eq:4}
		loss_{cla} = \frac{1}{R}\sum_{r=1}^{R}\sum_{q=1}^{Q} y_{rq}\log(p_{rq})
\end{equation}

Here, $R$ represents the number of training samples in the dataset, and $Q$ represents the number of categories for each sample ($Q \in {1,2,3}$). When $Q$ is equal to 1, 2, or 3, it represents the true carotid artery plaque category for that sample as hyperechoic plaque, hypoechoic plaque, or mixed-echoic plaque, respectively. For each input sample, the classification task network outputs a classification prediction $p_{r}$, where $p_{rq}$ represents the probability of type $q$ in the classification prediction, and $y_{rq}$ represents the value of the classification label.

\begin{equation}\label{eq:5}
		loss_{total} = loss_{cla}+loss_{seg}
\end{equation}

The total loss of WAL-Net is the sum of the classification loss $loss_{cla}$ and the segmentation loss $loss_{seg}$, as shown in Eq.\ref{eq:5}.

}
}

\section{Experiments and Results} 
{

\begin{figure}[!t]
	\centering
	\includegraphics[width=0.46\textwidth]{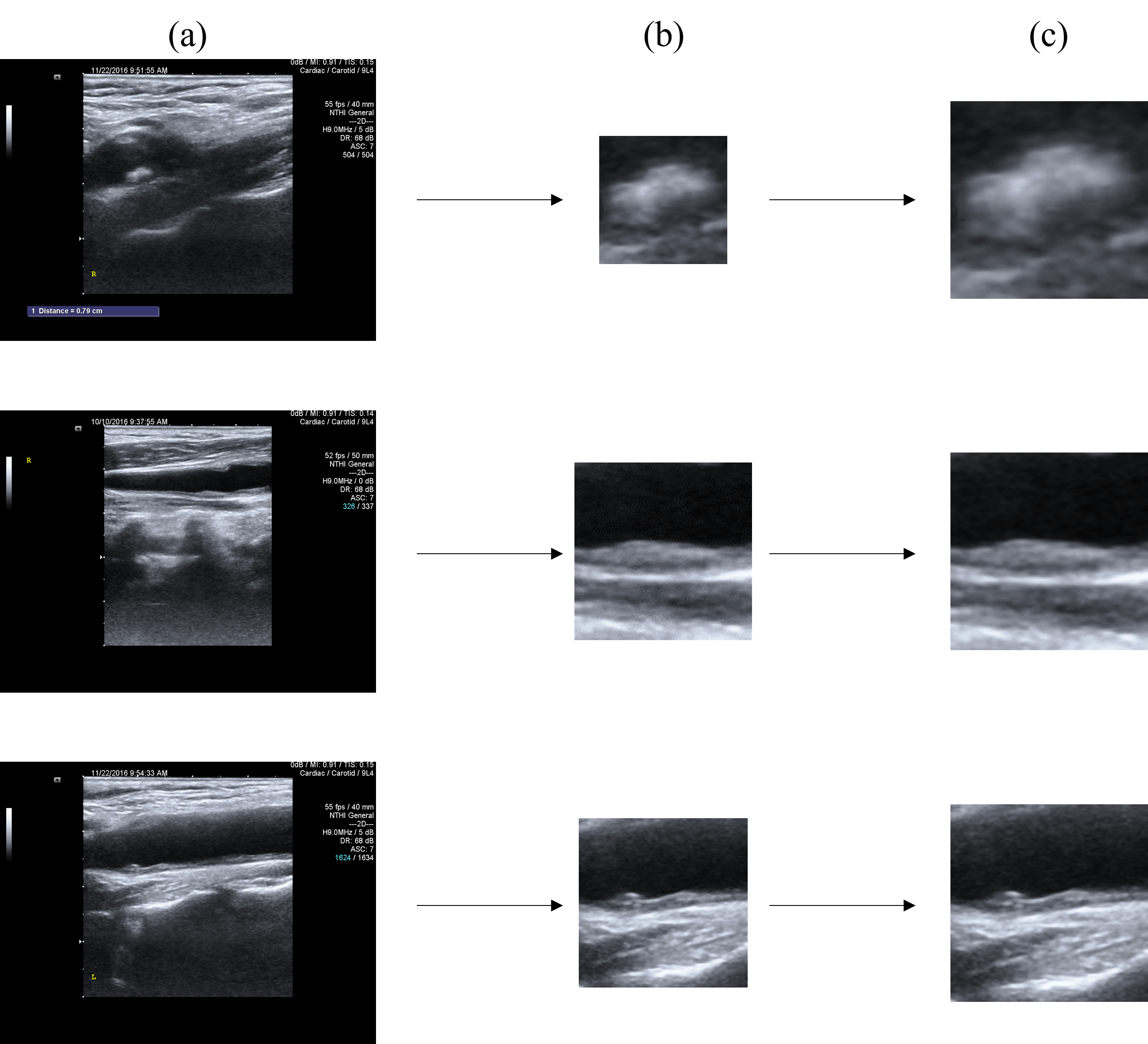}
	\caption{The preprocessing pipeline for carotid artery plaque ultrasound images. (a): The original ultrasound images of the three types of plaques: hyperechoic plaque, hypoechoic plaque, and mixed-echoic plaque. (b): The regions of interest in the ultrasound images of the three types of plaques. (c): The ultrasound images of the three types of plaques after being uniformly resized.}
	\label{fig:3}
\end{figure}

The data used for evaluating experiments in this paper is derived from a carotid artery ultrasound image dataset obtained from Zhongnan Hospital of Wuhan University. The study received approval from the Institutional Review Board (IRB) of the hospital. Each lesion image in the dataset is accompanied by a corresponding lesion category label. The dataset comprises a total of 1,270 carotid artery ultrasound images collected from 844 patients by ultrasound imaging experts. Among these images, there are 301 hyperechoic plaque images, 605 hypoechoic plaque images, and 364 mixed-echoic plaque images. For evaluation and experimentation purposes, the dataset is split into training, validation, and test sets in a 6:2:2 ratio. The code used in the experimental section of this paper has been uploaded to 'https://github.com/a610lab/WAL-Net'

\subsection{Data Preprocessing}
Due to variations in the sizes of the regions of interest in each sample of carotid artery ultrasound images (e.g., the smallest sample size is 19x29, while the largest is 134x564), this study standardizes all sample sizes to a uniform dimension (224x224, as used in the experiments). The following preprocessing steps are applied to generate input samples for the network model from the dataset: (1) Extract the region where the provided plaque is located in the carotid artery plaque ultrasound image, obtaining a rough region of interest image for each plaque as provided by the medical experts.
(2) Normalize the images obtained in step 1 to a consistent size of 224x224.The preprocessing steps are illustrated in Fig.\ref{fig:3}.

\subsection{Experimental Setup}
The experiments conducted in this study were implemented using the Python programming language and the PyTorch framework. The optimizer used in the experiments was the Adam optimizer with a learning rate of 0.0001. A batch size of 8 was selected for the experiments. All experimental results are the averages obtained after five random experiments on the same device.

Five evaluation metrics were defined for the classification of carotid artery plaque ultrasound images, including accuracy, F1-score, kappa, precision, and recall.

\subsection{Comparison with Other Classification Methods}
In our dataset, WAL-Net was compared with several state-of-the-art classification methods, including Convnext-v2[23], DPN[24], Repvit[25], Sequencer[26], Rexnet[27], Res2net[28], and Resnest[22]. The experimental results for these comparative methods were obtained by running their respective source codes.

\begin{table*}[!t]
	\centering
	\renewcommand{\arraystretch}{1.7} 
	\captionsetup{font=large}
	\caption{Performance comparison of classification methods on the carotid artery ultrasound image dataset. The best and the 2nd best results are marked in \textbf{bold} and $\underline{\mathrm{underline}}$.}
	\begin{tabularx}{\textwidth}{@{}l X X X X X@{}}
		\specialrule{1.2pt}{0pt}{0pt}
		\toprule
		\textbf{Method} & \textbf{Accuracy $\uparrow$} & \textbf{F1-score $\uparrow$} & \textbf{Kappa $\uparrow$} & \textbf{Precision $\uparrow$}  & \textbf{Recall $\uparrow$} \\ 
		\specialrule{0.6pt}{0pt}{0pt}
		\midrule
		ConvNext-V2(2023)  ~\citep{woo2023convnext}  & 0.6149 (0.065) & 0.6073 (0.162) & 0.3884 (0.162) & 0.6262 (0.194) & 0.6001 (0.111) \\ 
		DPN(2020) ~\citep{chen2017dual}      & 0.7762 (0.021) & 0.7665 (0.023) & 0.6378 (0.035) & 0.7993 (0.027) & 0.7536 (0.024) \\ 
		RepVit(2023)  ~\citep{wang2023repvit}  & 0.7042 (0.019) & 0.6927 (0.017) & 0.5226 (0.027) & 0.7205 (0.021) & 0.6863 (0.013) \\ 
		Sequencer (2022) ~\citep{tatsunami2022sequencer} & 0.6728 (0.022) & 0.6702 (0.029) & 0.4786 (0.038) & 0.7004 (0.022) & 0.6561 (0.032) \\
		RexNet (2020) ~\citep{han2020rexnet}  & 0.6444 (0.020) & 0.6370 (0.026) & 0.4346 (0.036) & 0.6498 (0.023) &0.6416 (0.029) \\ 
		Res2Net(2019) ~\citep{gao2019res2net}  & 0.8046 (0.018) & 0.8002 (0.016) & 0.6860 (0.029) & 0.8245 (0.024) & 0.7899 (0.017) \\ 
		ResNeSt(2022) ~\citep{zhang2022resnest}      & $\underline{0.8513}$ (0.018)  & $\underline{0.8473}$ (0.018) & $\underline{0.7641}$ (0.028) & $\underline{0.8570}$ (0.017) & $\underline{0.8437}$ (0.018) \\ 
		\textbf{WAL-Net (Ours)} & \textbf{0.8644} (0.011) & \textbf{0.8597} (0.011) & \textbf{0.7856} (0.016) & \textbf{0.8671} (0.017) & \textbf{0.8574} (0.009) \\ 
		\bottomrule
		\specialrule{1.2pt}{0pt}{0pt}
		\end{tabularx}
		\label{tab：1}
\end{table*}

Table 1 presents the comparative experimental results for the classification networks. WAL-Net achieved the best performance. Specifically, compared to the second-best performer Resnest (which is also the backbone network used in WAL-Net), WAL-Net demonstrated an improvement of approximately 1.3\% in the accuracy metric, highlighting the effectiveness of WAL-Net.

\begin{figure*}[htbp]
    \centering
    \subfigure[hyperechoic ROC]{
        \includegraphics[width=0.48\textwidth]{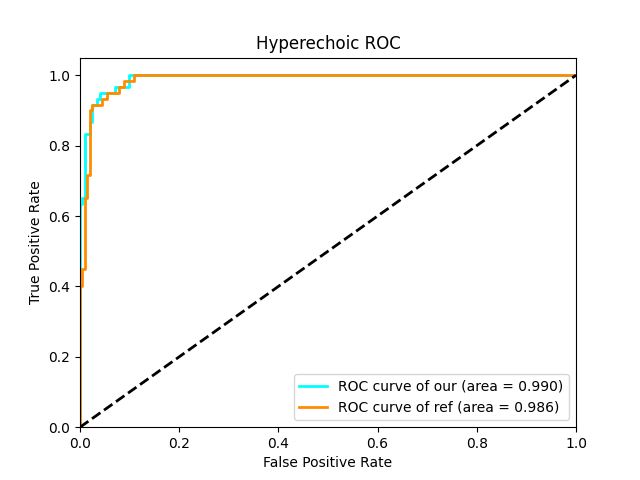}
    }
    \subfigure[hypoechoic ROC]{
        \includegraphics[width=0.48\textwidth]{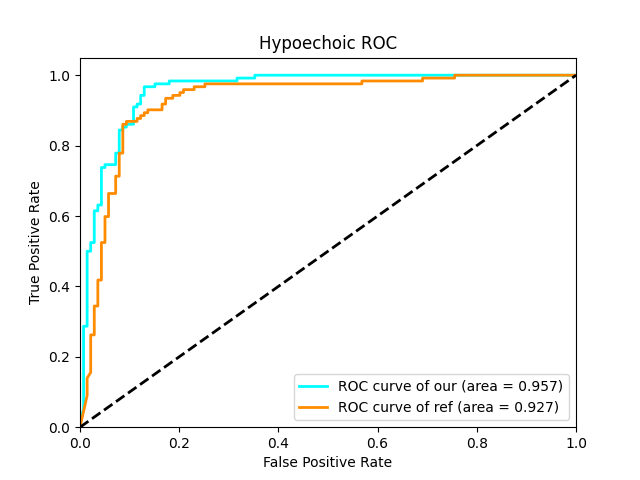}
    }
    \subfigure[mixed-echoic ROC]{
        \includegraphics[width=0.48\textwidth]{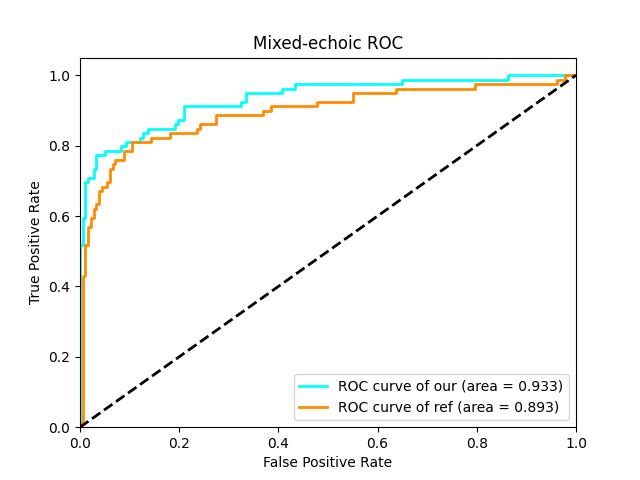}
    }
    \subfigure[micro ROC]{
        \includegraphics[width=0.48\textwidth]{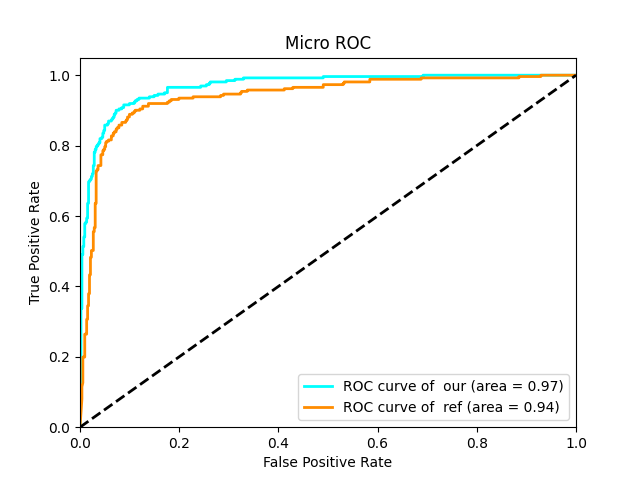}
    }
    \caption{ROC curves for WAL-Net and baseline.}
    \label{fig:4}
\end{figure*}

Our proposed WAL-Net demonstrates superior performance compared to the baseline network (Resnest[22]) on the dataset of carotid artery plaque ultrasound images. The ROC curves in Fig.\ref{fig:4} illustrate the performance of WAL-Net and the baseline network in classifying images from different categories in the dataset. It can be observed from the figure that WAL-Net outperforms the baseline network in all categories, especially in the recognition of hypoechoic plaques and mixed-echoic plaques. The performance difference between the two is not significant in the identification of hyperechoic plaques, possibly because this category inherently has a high recognition accuracy, with the ROC curve area exceeding 0.98. Overall, WAL-Net exhibits a noticeable improvement in performance compared to the baseline network, demonstrating the superiority of our approach.

\begin{figure*}[htbp]
    \centering
    \subfigure[Baseline confusion matrix]{
        \includegraphics[width=0.48\textwidth]{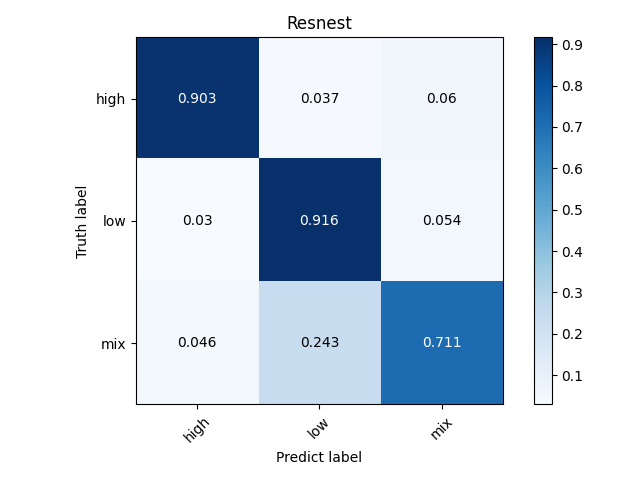}
    }
    \subfigure[WAL-Net confusion matrix]{
        \includegraphics[width=0.48\textwidth]{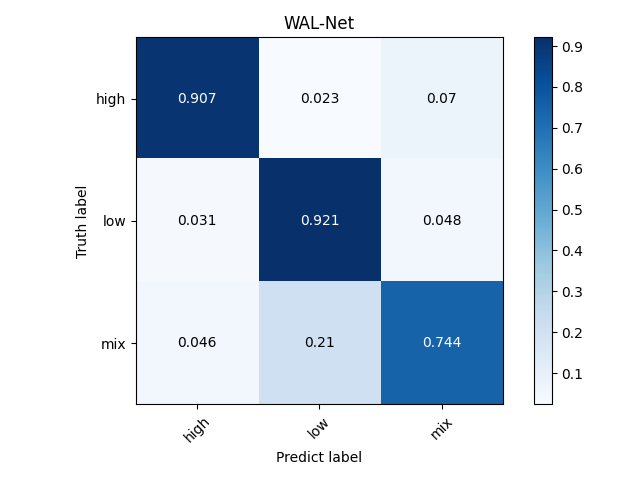}
    }
    \caption{Confusion matrix for WAL-Net and baseline.}
    \label{fig:5}
\end{figure*}

In carotid artery plaque ultrasound images, the convolutional network model faces varying levels of difficulty in learning between different categories. This discrepancy is partially due to the uneven distribution of samples among categories, differences in feature extraction difficulty for images from distinct categories, and other contributing factors.  Fig.\ref{fig:5} presents the confusion matrix of accuracy for different echo categories between WAL-Net and the baseline network. As observed in the figure, WAL-Net demonstrates a substantial improvement in recognition accuracy for hypoechoic plaques and mixed-echoic plaques, with an approximate increase of 3.3\% in accuracy for mixed-echoic plaques. The improvement in accuracy is less pronounced for hyperechoic plaques. Overall, WAL-Net achieves increased prediction accuracy across all three categories compared to the baseline.

\subsection{Visualization of Weakly Supervised Segmentation Results}
In Fig.\ref{fig:6}, we present visualization examples illustrating the pseudo-segmentation labels, segmentation predictions, and true segmentation labels obtained through our proposed weakly supervised segmentation method. As depicted in the figure, the segmentation predictions generated by our model effectively delineate plaque regions in ultrasound images, successfully suppressing noise originating from the vessel wall. This visualization highlights the model's ability to achieve accurate segmentation predictions without relying on true segmentation labels, showcasing its competitive performance.

\begin{figure}[!t]
	\centering
	\includegraphics[width=0.46\textwidth]{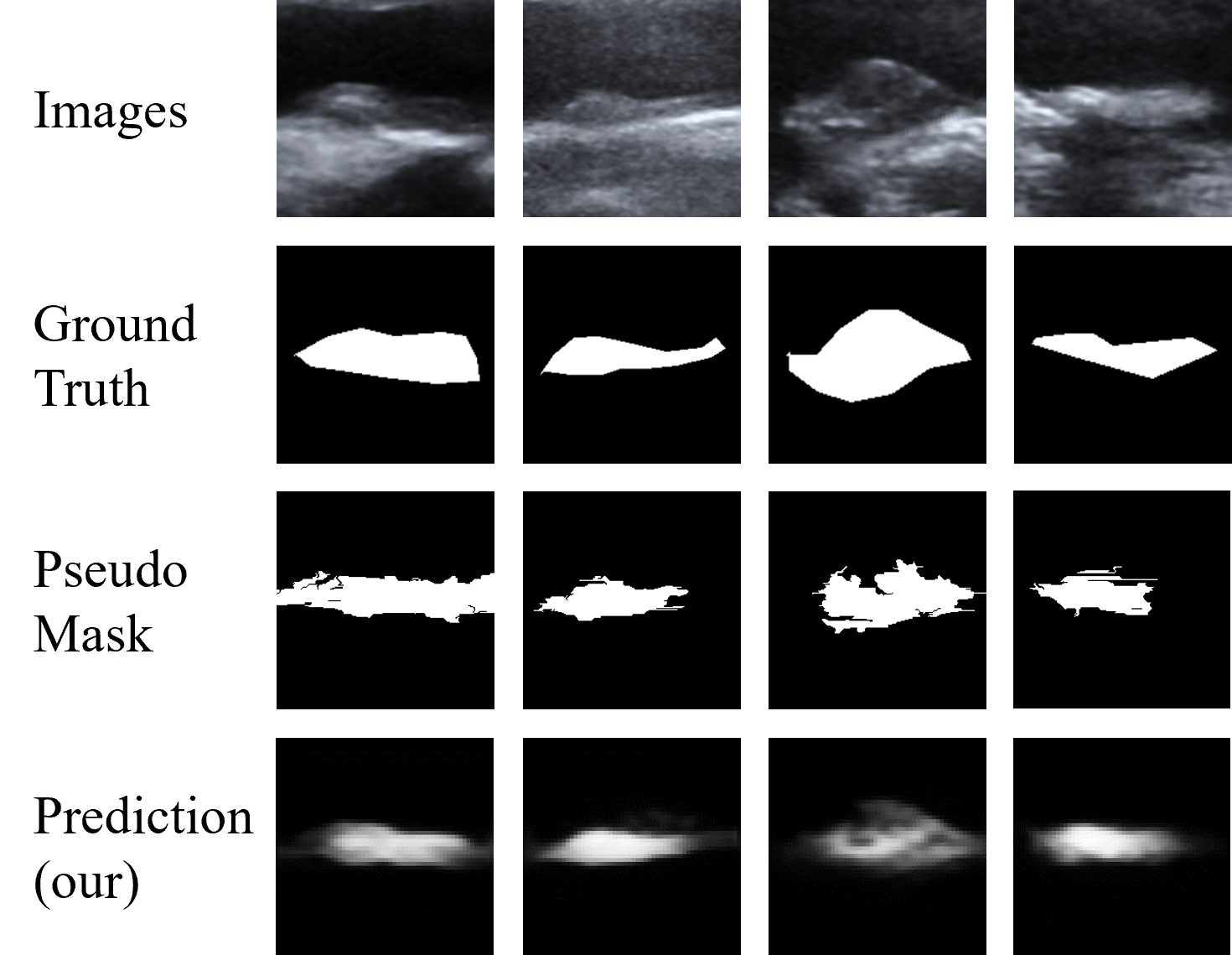}
	\caption{Visualization examples of the generated pseudo mask and the predicted results by proposed WAL-Net.}
	\label{fig:6}
\end{figure}

\subsection{Ablation Study}
In this table, we compare the experimental results of WAL-Net with different modules against the baseline of Resnest 50. The accuracy of the baseline Resnest 50 is 85.1\%. When incorporating an attention mechanism into the backbone network, the accuracy improves to 85.5\%. Furthermore, by adding the weakly supervised segmentation auxiliary task, WAL-Net achieves an accuracy of 86.4\%. These results demonstrate the effectiveness of both the attention mechanism and the weakly supervised segmentation auxiliary task in enhancing the recognition of carotid artery plaque ultrasound images.

\begin{table*}
	\centering
	\renewcommand{\arraystretch}{1.7} 
	\captionsetup{font=large}
	\caption{Ablation study for the RCM module and PGM module on the carotid artery ultrasound image dataset.}
	\vspace{-8pt}
	\begin{tabularx}{\linewidth}{@{}l X X X X X@{}}
		\specialrule{1.2pt}{0pt}{0pt}
		\toprule
		\textbf{Method} & \textbf{Accuracy $\uparrow$} & \textbf{F1-score $\uparrow$} & \textbf{Kappa $\uparrow$} & \textbf{Precision $\uparrow$}  & \textbf{Recall $\uparrow$} \\ 
		\specialrule{0.6pt}{0pt}{0pt}
		\midrule
		w/o RCM \& PGM & 0.8513 (0.018)  & 0.8473 (0.018) & 0.7641 (0.028) & 0.8570 (0.017) & 0.8437 (0.018) \\
		w/o RCM  & 0.8554 (0.015) & 0.8498 (0.016) & 0.7718 (0.024) & 0.8552 (0.014) & 0.8501 (0.014) \\ 
		WAL-Net & 0.8644 (0.011) & 0.8597 (0.011) & 0.7856 (0.016) & 0.8671 (0.017) & 0.8574 (0.009) \\  
		\bottomrule
		\specialrule{1.2pt}{0pt}{0pt}
	\end{tabularx}
	\label{tab：2}
\end{table*}

\subsection{Comparison with Different ROI Augmentation Methods}
In the work conducted by Amirreza Mahbod et al. ~\citep{mahbod2020effects}, various methods influencing segmentation for classification were experimentally explored, and a particularly effective non-end-to-end approach was identified. Extending this methodology to end-to-end networks, as demonstrated by He et al. ~\citep{he2023joint}, involved similar operations on the advanced features of classification networks, yielding favorable outcomes. To validate the applicability of this approach to the specific context of carotid artery ultrasound image datasets, we conducted experiments employing different segmentation strategies. As shown in Table 3, 'bg rm' represents the removal of background values from features, 'bg rm\&crop' represents eliminating background values and crop the foreground to a fixed size, 'crop' signifies no removal of background values but direct crop of the foreground to a fixed size, and 'rwm' refers to the method proposed by ~\citep{fu2023smc}, which multiplies the segmentation output predictions with advanced features for classification, emphasizing foreground weights while diminishing background weights. 'dilated crop' represents cropping the foreground and a portion of the surrounding background, then resizing to a fixed size, which is the method adopted in this paper. The results indicate the superiority of our proposed method over various strategies in the given context.

\begin{table*}
	\centering
	\renewcommand{\arraystretch}{1.7} 
	\captionsetup{font=large}
	\caption{Performance comparison of different ROI augmentation methods for RCM module.}
	\vspace{-8pt}
	\begin{tabularx}{\linewidth}{@{}l X X X X X@{}}
		\specialrule{1.2pt}{0pt}{0pt}
		\toprule
		\textbf{Method} & \textbf{Accuracy $\uparrow$} & \textbf{F1-score $\uparrow$} & \textbf{Kappa $\uparrow$} & \textbf{Precision $\uparrow$}  & \textbf{Recall $\uparrow$} \\ 
		\specialrule{0.6pt}{0pt}{0pt}
		\midrule
		rwm ~\citep{fu2023smc} & 0.8483 (0.011)  & 0.8438 (0.012) & 0.7593 (0.017) & 0.8526 (0.015) & 0.8394 (0.013) \\ 
		bg rm  & 0.8100 (0.020)  & 0.8025 (0.024) & 0.6932 (0.036) & 0.8335 (0.011) & 0.7884 (0.031) \\ 
		bg rm \& crop & 0.8352 (0.016) & 0.8305 (0.017) & 0.7397 (0.025) & 0.8358 (0.017) & 0.8297 (0.018) \\ 
		crop & 0.8452 (0.017)  & 0.8395 (0.018) & 0.7542 (0.029) & 0.8496 (0.015) & 0.8361 (0.021) \\ 
		\textbf{dilated crop} &\textbf{0.8644} (0.011) & \textbf{0.8597} (0.011) & \textbf{0.7856} (0.016) & \textbf{0.8671} (0.017) & \textbf{0.8574} (0.009) \\
		\bottomrule
		\specialrule{1.2pt}{0pt}{0pt}
	\end{tabularx}
	\label{tab：3}
\end{table*}

In Table 3, our adopted segmentation-influencing classification method demonstrates superior performance, outperforming the second-best rwm method by approximately 1.8\% in accuracy. This substantiates the effectiveness of our proposed approach.
}

\section{Conclusion}
\label{sec:conclusion}
In this paper, we posit that judiciously harnessing the intrinsic correlations between different tasks in auxiliary task learning is crucial for improving the classification results of carotid plaque ultrasound images. Consequently, we introduce a novel Weakly Supervised Auxiliary Task Learning Network (WAL-Net) comprising a shared encoder, a classification task head, and a weakly supervised segmentation task head. In contrast to traditional classification approaches, WAL-Net incorporates an auxiliary task based on weakly supervised learning, specifically, the segmentation task. Exploiting the auxiliary task, we explicitly enhance the classification task using the RCM module, thereby improving the performance of the classification task. We also design a module for the supervision of weakly supervised auxiliary task learning, utilizing the combination of unsupervised learning and attention mechanisms to generate pseudo-segmentation labels, thereby completely alleviating dependence on real segmentation labels while still achieving satisfactory segmentation results. Various experiments on the carotid ultrasound dataset demonstrate the effectiveness of our approach.

It is worth noting that optimizing the PGM module to obtain better pseudo-segmentation labels without relying on real labels will make the auxiliary task more effective and precise. In future work, we plan to expand the PGM module to enhance the performance of the auxiliary task. Furthermore, while our method is specifically applied to carotid plaque ultrasound images, WAL-Net and its individual sub-modules proposed herein are generalizable and hold potential for application in other image classification domains.

\section*{Conflict of interest statement}
There are no known conflicts of interest associated with this publication.

\section*{Authorship contribution statement}
\textbf{Haitao Gan}: Conceptualization, Methodology, Funding acquisition, Writing – original draft, Writing – Review \& editing. \textbf{Lingchao Fu}: Methodology, Software, Validation, Data curation, Writing – original draft. \textbf{Ran Zhou}: , Methodology, Funding acquisition, Validation, Writing – Review \& editing. \textbf{Weiyan Gan}: Investigation, Data curation. \textbf{Furong Wang}: Resources, Formal analysis, Data curation. \textbf{Xiaoyan Wu}: Resources, Formal analysis, Data curation. \textbf{Zhi Yang}: Conceptualization, Writing – Review \& editing. \textbf{Zhongwei Huang}: Conceptualization, Methodology.

\section*{Acknowledgment}
This work is supported by the High-level Talents Fund of Hubei University of Technology under grant No.GCRC2020016, Natural Science Foundation of China under grant No.62201203 and 62306106, Natural Science Foundation of Hubei Province under grant No.2023AFB377.

\bibliographystyle{plain}
\bibliography{refs}
\end{document}